\documentstyle[11pt,psfig,paspconf]{article}

\begin{document}

\thispagestyle{empty}
\vfill

\title{AGAPE: a Microlensing Search for Dark Matter by Monitoring Pixels}

\author{P.~Gondolo} 
\affil{Department of Physics, University of Oxford, 1
Keble Road, Oxford OX1 3NP, UK} 

\author{(For the AGAPE Collaboration)}

\vfill

\centerline{\it talk presented at the International Conference on}
\centerline{\it
Dark and Visible Matter in Galaxies, Sesto Pusteria, Italy, 2--5 July 1996}

\vfill\eject

\markboth{Gondolo, et al.}{AGAPE: a Microlensing Search...}
\setcounter{page}{1}

\title{AGAPE: a Microlensing Search for Dark Matter by Monitoring Pixels}

\author{P.~Gondolo} 
\affil{Department of Physics, University of Oxford, 1
Keble Road, Oxford OX1 3NP, UK} 

\author{For the AGAPE Collaboration: R.~Ansari,\altaffilmark{1}
M.~Auri{\`e}re,\altaffilmark{2} P.~Baillon,\altaffilmark{3}
A.~Bouquet,\altaffilmark{4} G.~Coupinot,\altaffilmark{2}
Ch.~Coutures,\altaffilmark{5} C.~Ghesqui{\`e}re,\altaffilmark{4}
Y.~Giraud-H{\'e}raud,\altaffilmark{4} J.~Hecquet,\altaffilmark{6}
J.~Kaplan,\altaffilmark{4} Y.~Le Du,\altaffilmark{4}
A.L.~Melchior,\altaffilmark{4} M.~Moniez,\altaffilmark{1}
J.P.~Picat,\altaffilmark{6} G.~Soucail\altaffilmark{6}}

\altaffiltext{1}{Laboratoire de l'Acc{\'e}l{\'e}rateur Lin{\'e}aire,
Universit{\'e} Paris-Sud, 91405 Orsay, France}
\altaffiltext{2}{Observatoire Midi-Pyr{\'e}n{\'e}es, unit{\'e}
associ{\'e}e au CNRS (UMR 5572), 62500 Bagn{\`e}res de Bigorre,
France} \altaffiltext{3}{CERN, 1211 Gen{\`e}ve 23, Switzerland}
\altaffiltext{4}{Laboratoire de Physique Corpusculaire, Coll{\`e}ge de
France, Laboratoire associ{\'e} au CNRS-IN2P3 (URA 6441),
11~place~Marcelin~Berthelot, 75231 Paris Cedex 05, France}
\altaffiltext{5}{SPP/DAPNIA, CEN Saclay, 91191 Gif-sur-Yvette, France}
\altaffiltext{6}{Observatoire Midi-Pyr{\'e}n{\'e}es, unit{\'e}
associ{\'e}e au CNRS (UMR 5572), 14 avenue Belin, 31400 Toulouse,
France }

\begin{abstract} 
AGAPE is an observational search of massive compact halo objects
(MACHOs) in the direction of M31 by means of a novel method: the
gravitational microlensing of unresolved stars. The search consists in
examining CCD pixel light curves for microlensing features. The high
level of temporal stability necessary to detect microlensing events
has been achieved, with quiet pixels stable within a factor of two of
the photon noise (the brightest ones down to a level of 0.001
mag). The data analysis is still in progress, but hundreds of variable
objects (cepheids, novae, ...) have already been found. Among them
there are several lightcurves that resemble microlensing events.
\end{abstract}


Dark matter could be in the form of compact baryonic objects
(MACHOs). A compact halo object passing in front of a star can act as
a gravitational lens and amplify the star light according to a
characteristic pattern in time. AGAPE is an observational search of
compact baryonic halo objects by means of a novel method: the
gravitational microlensing of unresolved stars (Baillon et al.\ 1992,
Crotts 1992). This amounts to monitoring the surface brightness of
millions of small zones on a distant galaxy. The main advantage with
respect to monitoring resolved stars is in the possibility to extend
microlensing searches to sources further away than the Galactic Center
and the Magellanic Clouds. AGAPE took CCD observations of the central
regions of the Andromeda galaxy (M31) at the 2-meter Telescope Bernard
Lyot (TBL) at Pic du Midi Observatory (France). The microlensing
search consists in examining pixel light curves for microlensing
features. In the reconstruction and calibration of the five million
pixel light curves, a high level of temporal stability has been
achieved, with quiet pixels stable within a factor of two of the
photon noise, sufficient for the detection of microlensing
events. These might be due either to Milky Way or Andromeda MACHOs or
to Andromeda bulge stars. Detection of the latter type of events would
validate the pixel method and would perhaps allow its absolute
calibration. The last observations were taken at the end of 1995, and
the data analysis is still in progress.  Hundreds of variable objects
have already been found. Some of them have been identified as Novae
and Cepheids, others await identification. Among them there are
several lightcurves that resemble microlensing events, but at the
moment we are unable to tell their nature.  Future AGAPE observations
at Pic du Midi in Fall 1996 are expected to provide useful
information. A complete account of observations and data analysis,
with full references, is given in Ansari et al.\ 1996.

\section{Microlensing of unresolved stars}

Gravitational microlensing as a means of searching for dark matter
objects has been proposed by Paczy\'nski in 1986. Two groups have
successfully searched in the direction of the Large and Small
Magellanic Clouds: the MACHO collaboration (Alcock et al.\ 1993,
1995a, 1996) and the EROS collaboration (Aubourg et al.\ 1993, Ansari
et al.\ 1995). And clear evidence of microlensing has been observed
towards the Galactic Center by the OGLE (Udalski et al.\ 1993, 1994),
MACHO (Alcock et al.\ 1995b) and DUO (Alard et al. 1995, Alard, Mao \&
Gilbert 1995) collaborations. To extend the microlensing search to
other lines of sight, so that additional information be available on
the properties of the lensing population, demands monitoring stars in
external galaxies in regions where stars are not resolved. A field of
view encompassing most of the external galaxy would also allow the
exploration of the dark halo associated with the external galaxy
itself. An unresolved star needs to be amplified enough by
microlensing so that the increase in luminosity in a resolution
element is detectable above the noise fluctuations of the background
from the other unresolved stars. This is expected to occur at a rate
of $\sim 10$ per year in the central regions of M31 for 0.08$M_{\sun}$
lenses belonging to a standard galactic dark halo (Baillon et al.\
1993, Ansari et al. 1996).

The shape of the resolution element lightcurve does not differ from
that of a single lensed star but for the constant baseline being the
integrated starlight in the resolution element instead of the light
$F_{*}$ from the lensed star only. So the flux variation
\begin{equation}
\Delta F_{\rm res.el.}(t) = \Delta F_{*}(t) = F_{*} \left[ A(t)-1
\right] ,
\end{equation}
where $A(t)$ is the microlensing amplification. Typically, it will not
be possible to measure $F_{*}$, but many of the tests for microlensing
of resolved stars can be applied to discriminate against light
variations of other origin: non-repetition, time symmetry,
achromaticity, spatial distribution. For achromaticity, one rather
uses the ratio of flux variations in two colors instead of the ratio
of amplifications.

In CCD observations from the ground, light is collected not in
resolution elements but in CCD pixels, and the background includes not
only the constant surface brightness of the target galaxy but also the
variable sky brightness, and it further varies according to atmospheric
absorption. A high-precision photometric alignment of the CCD images
(better than $1\%$) is needed. The imperfect telescope pointing
requires translation/rotation/distortion of the CCD images to
superpose the same surface brightness elements on the target galaxy
(geometric alignment). The way AGAPE addresses these observational
challenges is described next.

\section{Observations and data reduction}

We took data on the 2-meter telescope TBL during 76 nights of good
weather scattered over 2 months in 1994 (Sept 28--Nov 24) and 6 months
in 1995 (July--Dec). Observations were only carried out when M31 was
higher than 35\deg\ above the horizon. We used two well-separated
filters, Johnson B and Gunn r, with exposure times of 30 min and 20
min respectively. We observe at the f/25 Cassegrain focus behind a
focal reducer ``ISARD'' that brings the aperture to f/8. The camera is
a $1024\times 1024$ Tektronix CCD with 24 $\mu$m (0.3 arcsec)
pixels. The effective field of view covered by ISARD is $900\times
870$ pixels. We took images of 6 adjacent fields (A--F) in a $3\times
2$ pattern centered on M31. 

For various reasons, and in particular because the telescope is not
dedicated, the focal reducer ISARD must often be dismounted and
remounted. After such an operation, the positions of the mirrors and
the camera are never exactly the same as before. Flat-fields correct
for these differences, except for those arising from an ISARD retuning
in October 1994 for which a more sophisticated procedure was adopted
(see below).

Geometric alignment is performed by matching foreground stars on the
current and reference image. (Reference images are those of 1994
October 26.) A general linear transformation (6 parameters)
is fitted to the star positions, and then applied to the current
image to align it with the reference. Since great care was used in
repointing the fields, relative rotations and distortions between
images are very small.  Although they are important for the position
of the transformed pixel, they induce negligible changes of pixel
orientation, size and shape, so that pixel intensities can be obtained
by bilinear interpolation, The success of this geometric alignment is
seen in fig.~\ref{geom-align},
\begin{figure}
\centerline{\psfig{file=fig06.ps,width=0.5\textwidth}}
\vspace{-.5cm}
\caption{Residuals of star positions after geometric alignment of two
images.} \label{geom-align}
\end{figure}
where the residuals of star positions after alignment are plotted. The
dispersion is of order 0.1 arcsec, compared with a typical seeing of
1--2 arcsec. 

Photometric alignment is obtained with an original method based on
matching the intensity and gradient of the surface brightness of M31
across images. Indeed, due to the small number of foreground stars and
their poor photometry in the presence of the steep gradient, the
traditional method of comparing their fluxes is not precise enough for
this study. Since the luminosity gradient of the bulge of M31 is much
greater than the intensity fluctuations due to photon noise and seeing
variations, we safely assume a linear relation between the intensities
of corresponding pixels on the current and reference images. The
coefficients of this intensity transformation, giving the relative
absorption and the difference in sky backgrounds, are obtained from
the averages and standard deviations of the pixel histograms of the
two images. The efficiency of this procedure is illustrated in
fig.~\ref{photo-align}:
\begin{figure}
\vspace{1.5cm}
\begin{center} \hskip .25\textwidth (a) \hskip .5\textwidth (b) \end{center}
\vspace{-1.5cm}
\centerline{\psfig{file=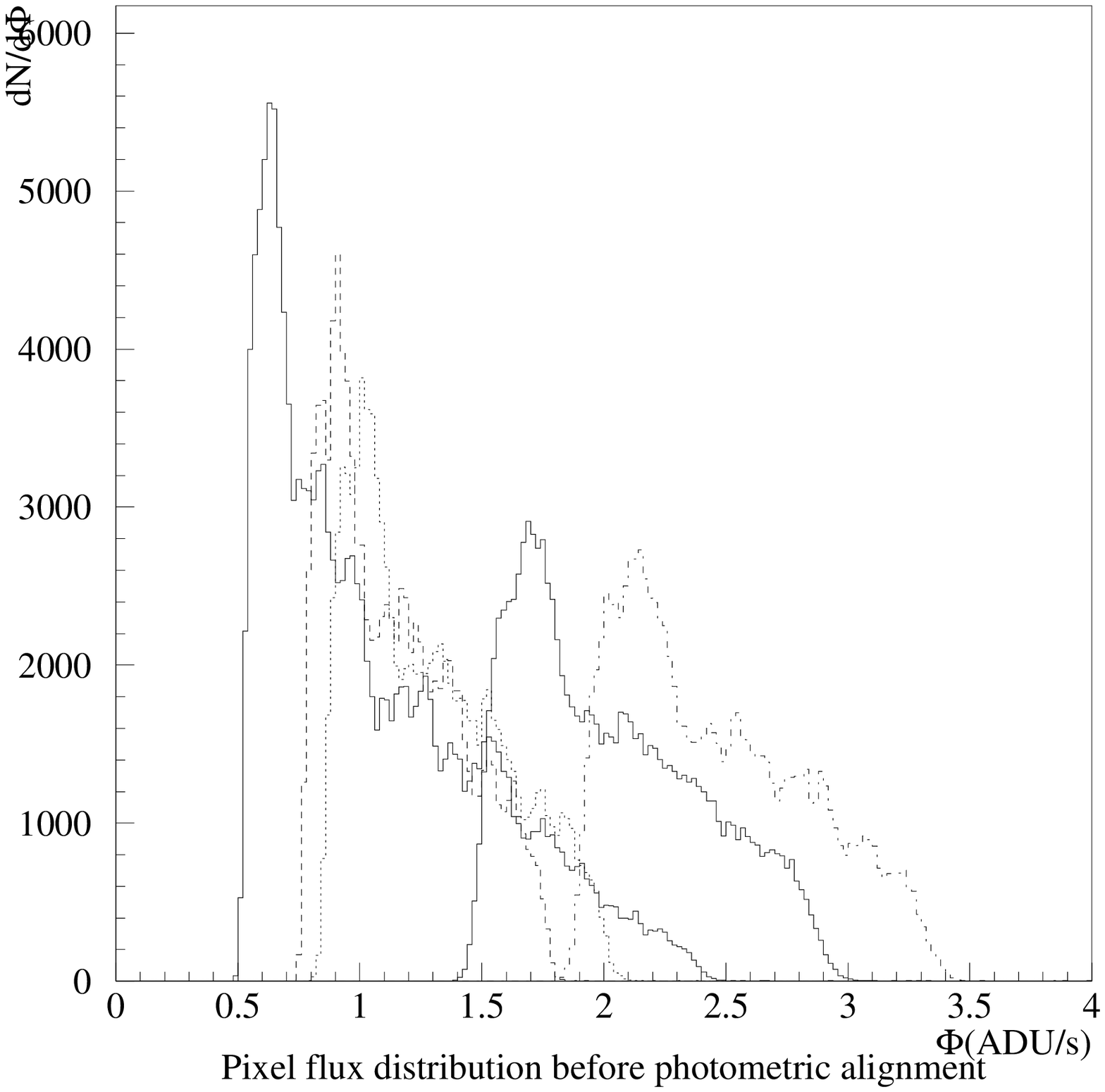,width=.48\textwidth}
\psfig{file=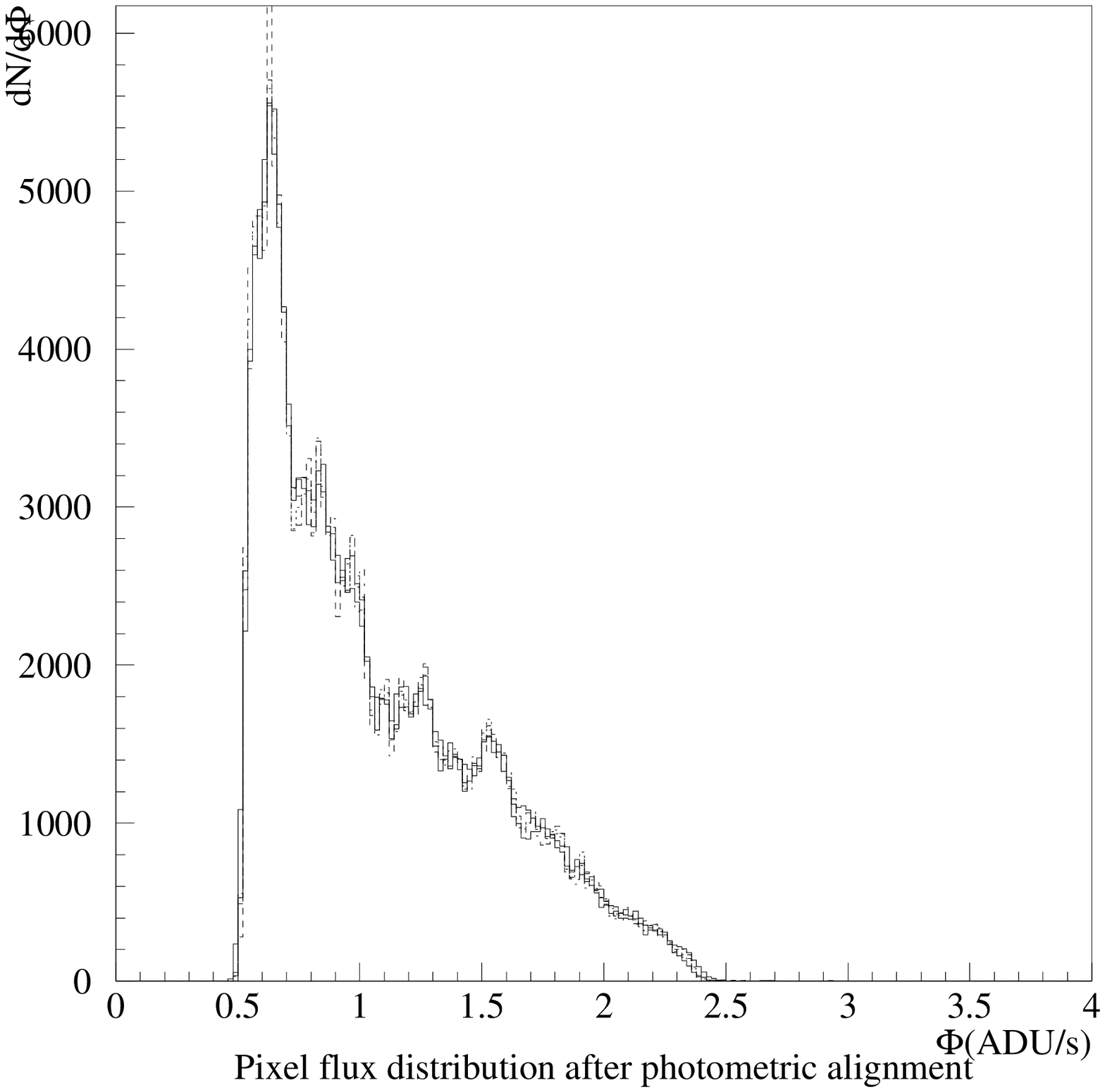,width=.48\textwidth}}
\vspace{-.3cm}
\caption{Histograms of pixel intensities (a) before and (b) after photometric 
alignment.} \label{photo-align}
\end{figure}
histograms of pixel intensities for five images coincide after
photometric alignment down to fine details. This procedure was
satisfactory for photometric alignment of most images, except between
those before and after the ISARD retuning for which preliminary
application of a high-pass spatial filter (median filter) was required
(see Ansari et al. 1996 for details).

Seeing variations induce unwanted fluctuations of the individual pixel
fluxes. To reduce them, an integration over the point spread function
should be performed. For the moment, we replace the intensity of each
individual pixel by the sum of the pixel intensities of a $7\times 7$
square centered on it, which we call a super-pixel. This 2.1 arcsec
wide integration zone is large enough to cope with seeing variations
when the seeing is below $1.8$ arcsec. Better integration methods
are under study.

\begin{figure}
\centerline{\psfig{file=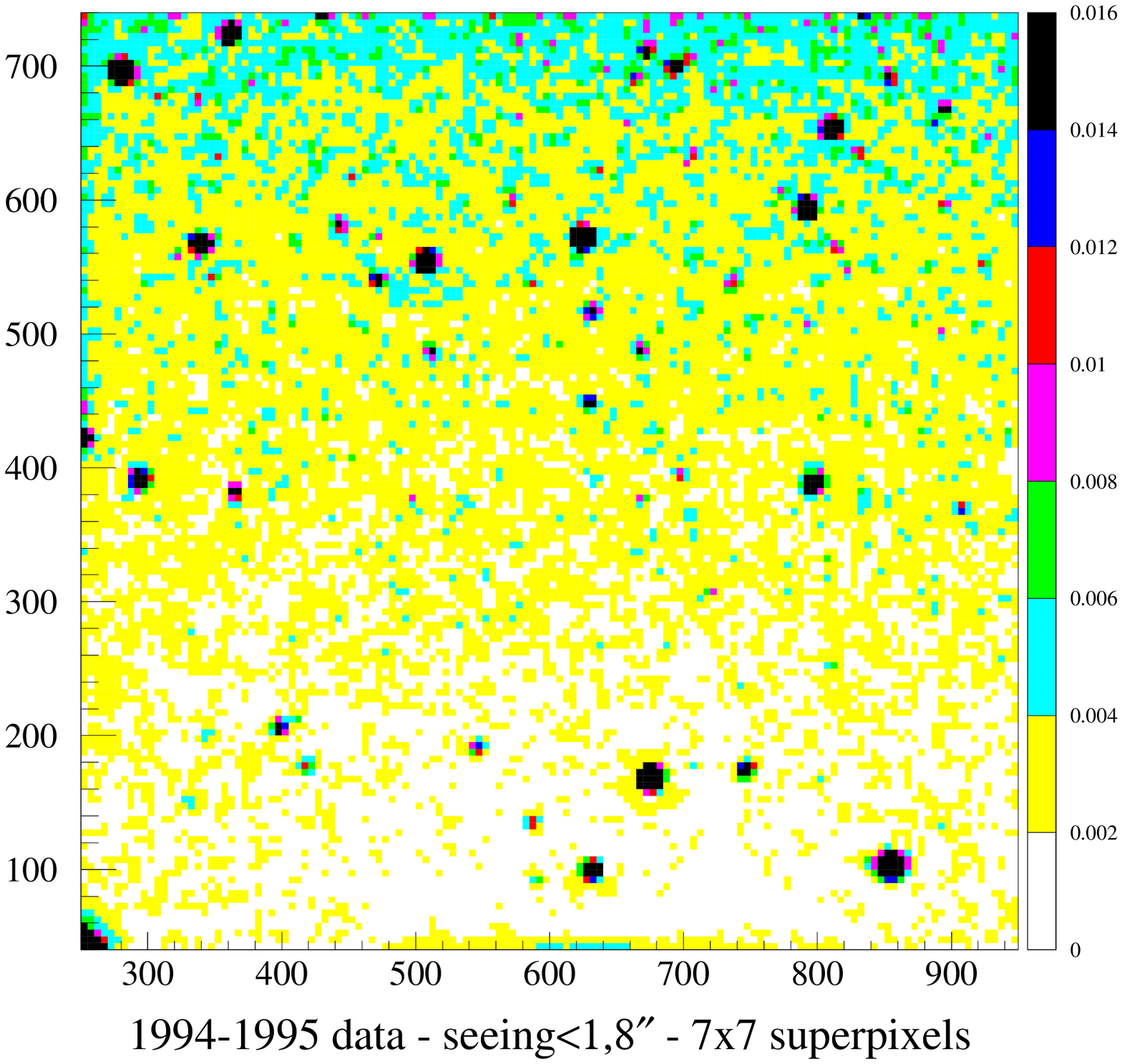,width=0.7\textwidth}}
\caption{Relative r.m.s.\ fluctuation of field-A super-pixel
lightcurves.} \label{rms-spix}
\end{figure}

\begin{figure}
\centerline{\psfig{file=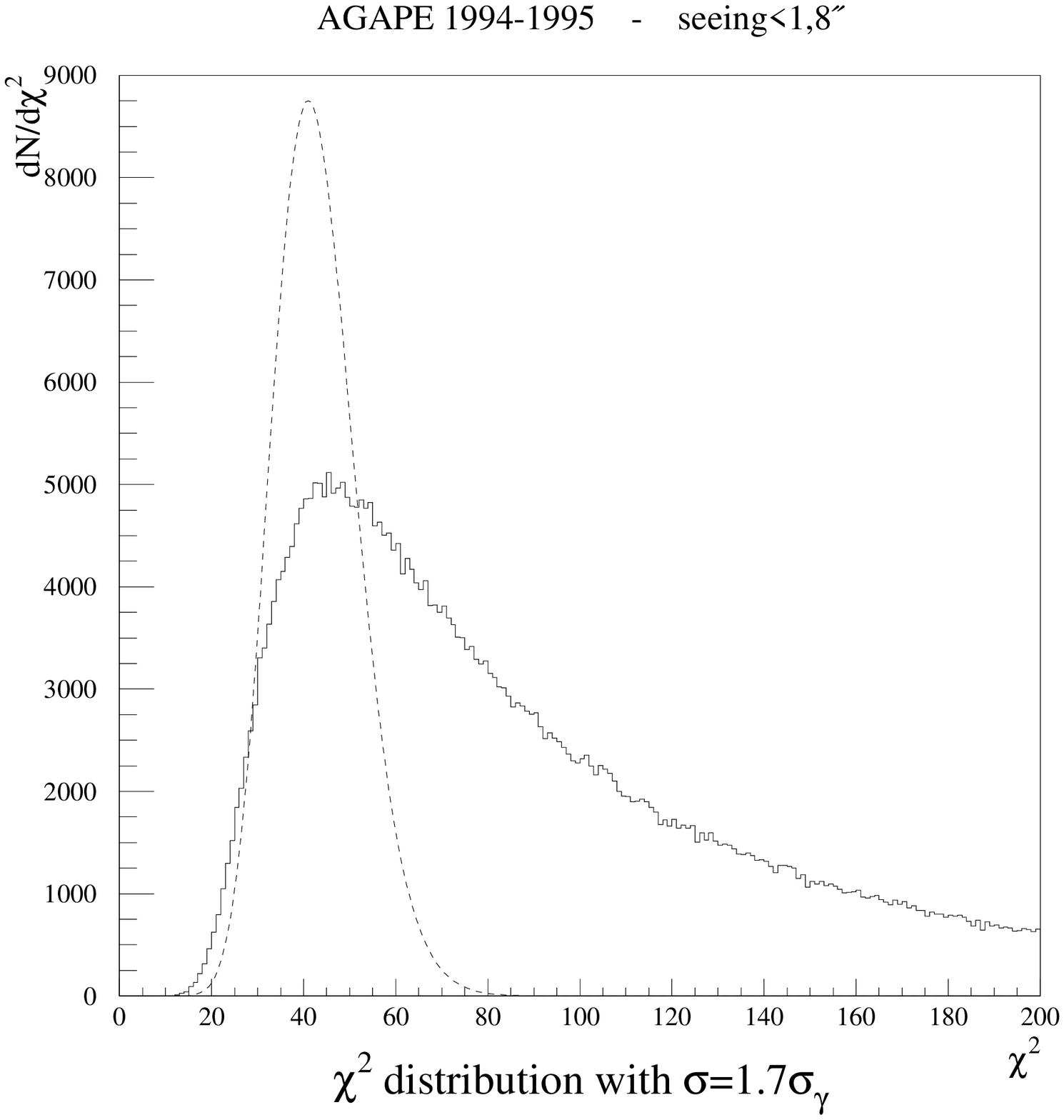,width=0.48\textwidth}}
\caption{$\chi^2$-distribution for field-A super-pixel lightcurves
(solid line) compared with the $\chi^2$-distribution from Gaussian
errors whose dispersion equals 1.7 times the Poissonian photon noise
(dotted line). Only images with seeing smaller than 1.8 arcsec are
included. } \label{chi2}
\end{figure}

The time stability of each super-pixel can be quantified by the
relative r.m.s.\ fluctuation of its intensity around its time
average. The map in fig.~\ref{rms-spix} 
shows this relative dispersion for super-pixels of field A. Dark spots
correspond to foreground stars and should be disregarded. The
fluctuation is typically of 0.3\% and can be as low as 0.1\% in the
most stable regions (next to the M31 center). To compare it to the
photon counting noise, we have computed the $\chi^2$ distribution of
each super-pixel lightcurve using errors proportional to the
Poissonian photon noise, and have found that a constant of
proportionality equal to 1.7 makes the maxima of the computed and
Poissonian distributions coincide (fig.~\ref{chi2}). 
The observed distribution shows however non-poissonian tails, probably related
to seeing variations.

Globally, we can say that geometric alignment of images is not a
serious problem, and that our photometric alignment produces
lightcurves of quiet super-pixels that are stable to around twice the
photon noise.  We are now in a position to select variable super-pixel
lightcurves, and to discriminate those corresponding to microlensing
from those due to variable sources.

\section{Super-pixel lightcurves}

The outcome of the previous data reduction are super-pixel
lightcurves. We present here examples of such lightcurves in the Gunn
r filter, as illustration of the possibilities of our method.

\begin{figure}
\centerline{\psfig{file=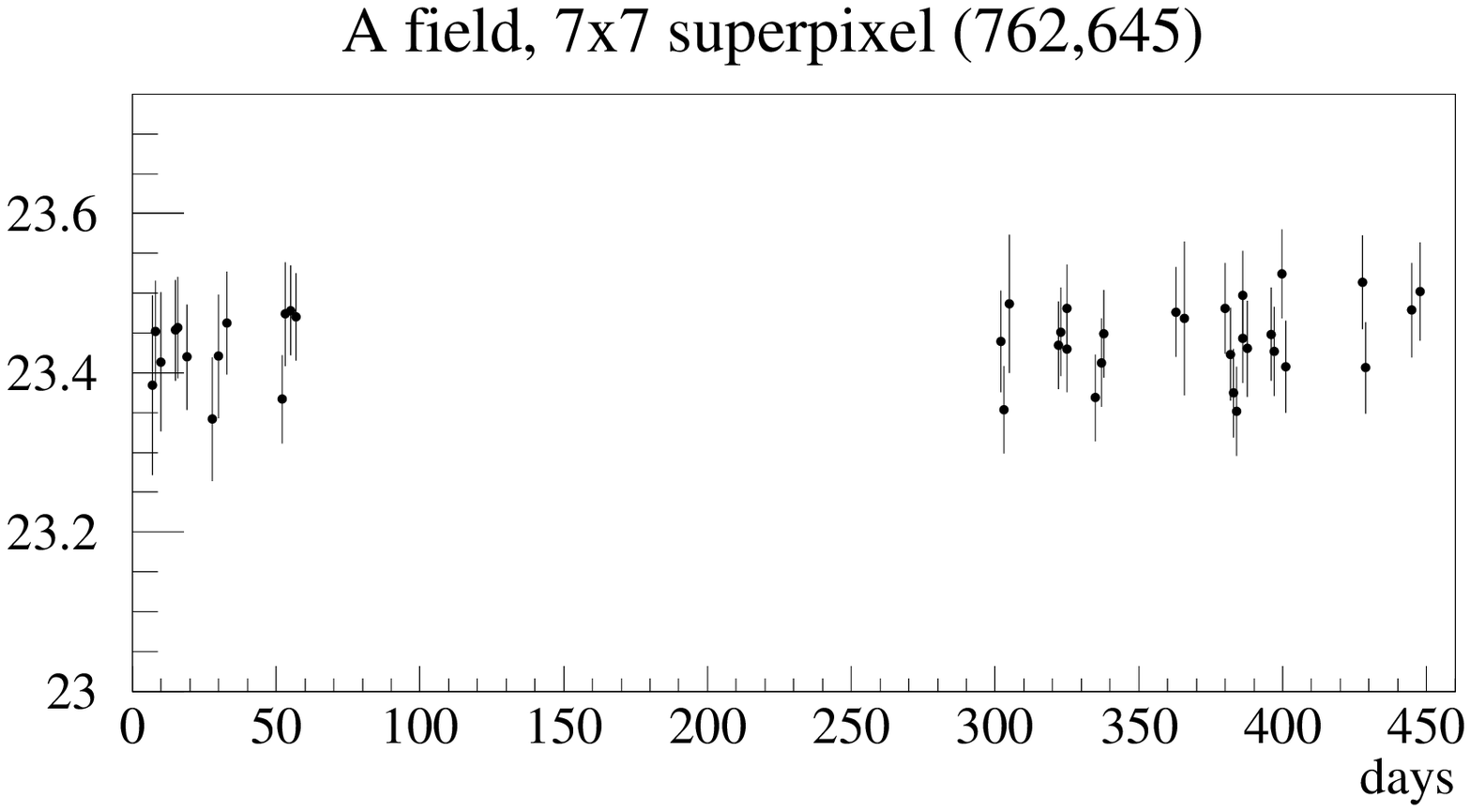,width=0.7\textwidth}}
\vspace{-.3cm}
\caption{The lightcurve of a typical quiet super-pixel. Fluctuations
are at a level of 0.002 mag. The vertical scale is in ADU/s.}
\label{quiet}
\vspace{.3cm}
\end{figure}

Fig.~\ref{quiet} shows the lightcurve of a quiet super-pixel.
The vertical axis is in ADU/s (1 ADU/s in Gunn r corresponds to a
surface brightness $\mu_r=22.1$).  The error bars are set at 1.7 times
the photon counting noise, where the factor 1.7 arises as argued
above. The r.m.s.\ fluctuation of this lightcurve is 0.045 ADU/s,
smaller than the average of the error bars (0.065 ADU/s). Note that
this corresponds to 0.002 mag. Super-pixels are indeed very stable and
error bars at 1.7 the photon noise are realistic. This is true for
most super-pixels in our fields.

\begin{figure}
\centerline{\psfig{file=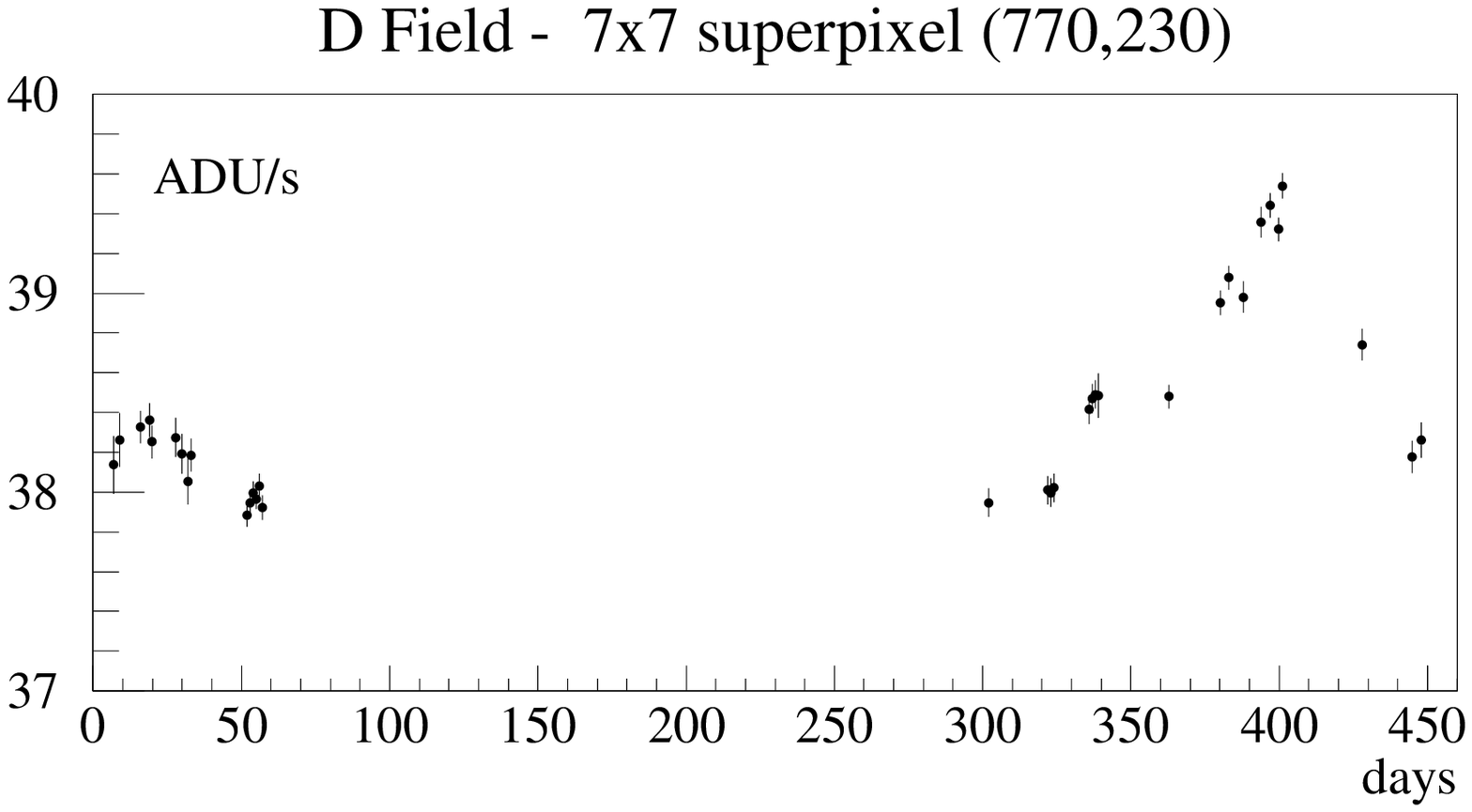,width=0.7\textwidth}}
\vspace{-.3cm}
\caption{The lightcurve of a variable super-pixel, showing a tiny but
detectable variation of 0.01 mag from day 20 to day 60. Only images
with seeing between 1.1 and 1.8 arcsec are shown.} \label{tiny}
\vspace{.3cm}
\end{figure}

We are able to clearly see variations at a level of a few per cent, as
is apparent in fig.~\ref{tiny}.
The first variation, from day 20 to day 60, is visible because of its
coherence in time. It is of only 0.01 mag, but this is about 5 times
the average error bar in this period.

Hundreds of variable objects like the preceding one have been
detected, and we are only beginning to analyse their nature. We have a
host of cepheid candidates and five novae, with peak magnitude and
rate of decrease similar to the M31 novae quoted in Hodge 1992.

\begin{figure}
\centerline{\psfig{file=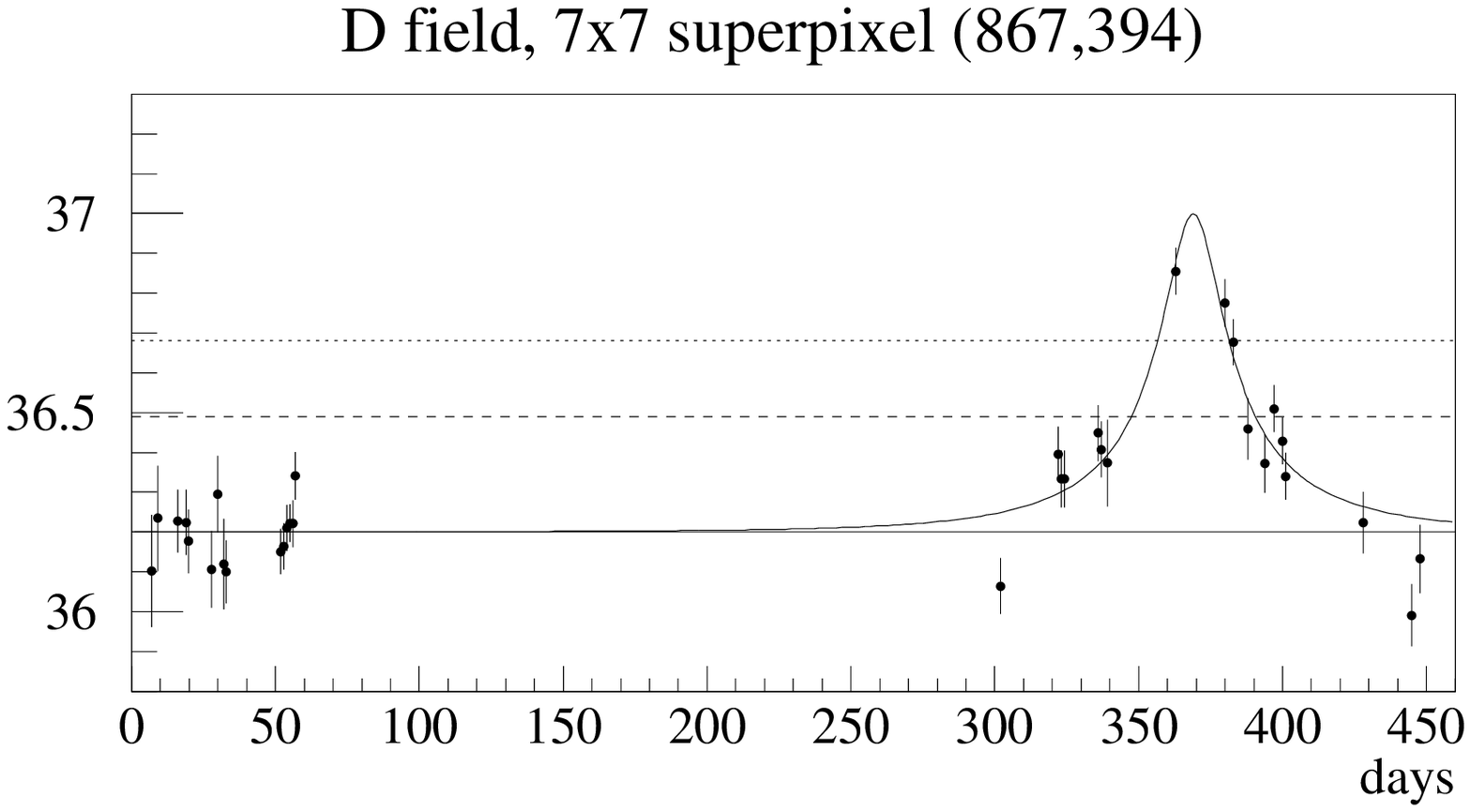,clip=,width=0.7\textwidth}}
\vspace{-.3cm}
\caption{Super-pixel lightcurve resembling a microlensing event. The
solid line is a fit of the theoretical microlensing curve. The solid
horizontal line is the baseline, the dashed line lies 3$\sigma$ and
the dotted line 5$\sigma$ above the baseline. Vertical scale is in
ADU/s.} \label{micro}
\vspace{.3cm}
\end{figure}

We also see variations compatible with microlensing events, but at this
stage we are not in a position to claim that they indeed are microlensing
events. This because (i) our coverage in time is too short to be sure
that the variations do not repeat or that some of the events are
really symmetric, and (ii) we have not yet analysed the blue
lightcurves and so cannot test achromaticity. Fig.~\ref{micro}
shows one of the lightcurves that resemble microlensing events. The
Paczy\'nski curve corresponds to a star of absolute magnitude $M=-2$
amplified by a factor of 6 at maximum and to an Einstein time scale
$t_E=65$ days. Because of a degeneracy for high amplification events
(see e.g. Gould 1995), a time scale and maximum amplification twice as
large associated with a star twice fainter would fit just as
well. However the time scale cannot be much shorter than 65 days,
because the star should have been brighter and would have been seen
before being lensed. 

Our simulations for galactic lenses of mass 0.08$M_{\sun}$ (100\% of
standard halo) predict that at our level of stability we should detect
$\sim 10$ microlensing events in our data, an event being detectable
if its lightcurve remains 3$\sigma$ above the background in at least 3
consecutive images and reaches 5$\sigma$ in at least one of them. We
have shown here that such variations are clearly detectable from their
time coherence, but more work is still needed to discriminate
microlensing events from other kinds of variations.

\acknowledgements

We are grateful to F.~Colas, D.~Gillieron, and A.~Gould for useful
discussions and suggestions.

\end{document}